\documentclass[journal=nalefd]{achemso}

\usepackage{amsmath}
\usepackage{graphicx}
\usepackage{tabularx}
\usepackage{hyperref}
\usepackage[table,xcdraw]{xcolor}
\usepackage{chemformula}

\usepackage[normalem]{ulem}

\usepackage{soul}

\usepackage{ulem,xpatch}

\xpatchcmd{\sout}
  {\bgroup}
  {\bgroup}
  {}{}
  
\author{D. Soriano}
\affiliation{Radboud University, Institute for Molecules and Materials, NL-6525 AJ Nijmegen, the Netherlands}
\email{d.soriano@science.ru.nl}
\author{M. I. Katsnelson}
\affiliation{Radboud University, Institute for Molecules and Materials, NL-6525 AJ Nijmegen, the Netherlands}
\author{J. Fern\'andez-Rossier}  
\affiliation{QuantaLab, International Iberian Nanotechnology Laboratory (INL), Av. Mestre Jos\'e Veiga, 4715-330 Braga, Portugal}
\alsoaffiliation{Departamento de F{\'i}sica Aplicada, Universidad de Alicante, 03690, Spain}

\title{Magnetic Two-Dimensional Chromium Trihalides: A Theoretical Perspective}

\begin{document}

\begin{abstract}
The discovery of ferromagnetic order in monolayer 2D crystals has opened a new venue in the field of two dimensional (2D) materials. 2D magnets are not only interesting on their own, but their integration in van der Waals heterostructures allows for the observation of new and exotic  effects in the ultrathin limit. The family of Chromium trihalides, CrI$_3$, CrBr$_3$ and CrCl$_3$, is, so far, the most studied among magnetic 2D crystals. In this mini-review, we provide a perspective of the state of the art of the theoretical understanding of magnetic 2D trihalides, most of which will also be relevant for other 2D magnets, such as vanadium trihalides. We  discuss both the well-established facts, such as the origin of the magnetic moment and magnetic anisotropy and address as well open issues such as the nature of the anisotropic spin couplings and the magnitude of the magnon gap. Recent theoretical predictions on Moir\' e magnets and magnetic skyrmions are also discussed. Finally, we give some prospects about the future interest of these materials and possible device applications.
\end{abstract}

\maketitle


{\section{Introduction}} The interplay between dimensionality and magnetic order plays a central role in the theory of magnetism. Thus, the exact solutions of the Ising model, show that spontaneous magnetic order forbidden in 1D and only posible in 2D. The Mermin-Wagner theorem precludes the existence of magnetic order at finite temperature in two dimensional magnets with isotropic  long range exchange interaction. These rigorous results already highlight the relevance of the symmetry of spin interactions for the existence of magnetic order in 2D.  Before the advent of 2D crystals, the experimental study of these matters could only be tackled in layered magnetic materials with weak, or even negligible,  interlayer interactions.\cite{deJongh2012,McGuire2017}

The discovery of graphene and other 2D materials opened the gate to the discovery of magnetic 2D crystals. An early  theoretical suggestion of 2D ferromagnets was to use a single-layer K$_2$CuF$_4$\cite{SachsWehling2013} but, despite its low theoretically predicted interlayer cohesive energy, attempts to exfoliate a single layer of this material were unsuccessful. Ferromagnetic order, down to monolayer was first reported in CrI$_3$ \cite{HuangXu2017}, at the same time as the few layer ferromagnetism in CrGeTe$_3$\cite{GongZhang2017}, at sub liquid nitrogen temperatures.  These seminal works were soon followed by many others, including  insulators such as  CrCl$_3$, CrBr$_3$, VI$_3$,\cite{KongCava2019} and  conducting monolayers such as Fe$_3$GeTe$_2$,\cite{DengZhang2018} and CrTe$_2$,\cite{SunZhang2019} that shows magnetic order  at  room temperature and MnBi$_2$Te$_4$ that displays quantized anomalous Hall effect.\cite{DengZhang2020} 
  
Probing magnetic order down to the monolayer is beyond reach of conventional magnetometry for micron scale crystals. As a result, several probing strategies have been used, including most-notably magneto-optical Kerr effect\cite{GongZhang2017,HuangXu2017}, Raman\cite{ZhangHuang2020},  and second harmonic generation\cite{SunWu2019}.  Recently, NV-center magnetometry has also  been used \cite{ThielMaletinsky2019,Rossier2019}, which permits to scan magnetic order with a 100nm resolution. 

Although the electronic properties of these ferromagnetic 2D crystals are interesting and  intriguing  on their own right,  the properties of Van der Waals (VdW) structures that integrate magnetic 2D crystals  takes this whole field into a completely new dimension, given the many outstanding possibilities. For starters,  both theory and experiments show a rich interplay between stacking and interlayer exchange\cite{McGuireSales2015,SorianoRossier2019,SivadasXiao2018}, which eventually may lead to a very interesting and experimentally unexplored physics in twisted few-layer samples.\cite{HejaziBalents2020} Besides, the weak magnetic exchange interaction in antiferromagnetic bilayer CrI$_3$ allows for a field effect control of the magnetic properties of CrI$_3$ bilayers, as demonstrated both experimentally\cite{JiangMak2018,HuangXu2018} and theoretically.\cite{SorianoKatsnelson2020} Finally, the demonstration of magnon-magnon interactions via electron tunneling spectroscopy in CrI$_3$  \cite{KleinJarillo2018} and CrBr$_3$\cite{GhazaryanMisra2018}, and the observation of spin-filter tunnel magnetoresistance\cite{KleinJarillo2018,SongXu2018,WangMorpurgo2018} make these materials very attractive for the design of future devices based on van der Waals heterostructures.
 
Spin proximity effect between Cr trihalides and semiconducting transition metal dichalcogenides has been observed by means of magneto-optical probes by several groups\cite{ZhongXu2017,ZhongXu2020}. 
The observation of a zero bias peak at the edge of a CrBr$_3$ monolayer grown on top of a superconducting material has been interpreted in terms of the emergence of topological superconductivity and the concomitant Majorana particle.\cite{KezilebiekeLiljeroth2020}

In this review, we give a brief overview of the theoretical aspects regarding magnetism in two dimensional materials. We also report on the most recent theoretical advances, and open issues, in understanding the magnetic exchange interactions and anisotropic spin couplings of chromium trihalides. The recent observation of skyrmions in twisted few-layer samples, and the realization of topological magnons are also discussed. Finally, we give some prospects regarding the interest of these materials and future device applications.

\begin{figure*}[t]
	\centering
	\includegraphics[clip=true, width=\textwidth] {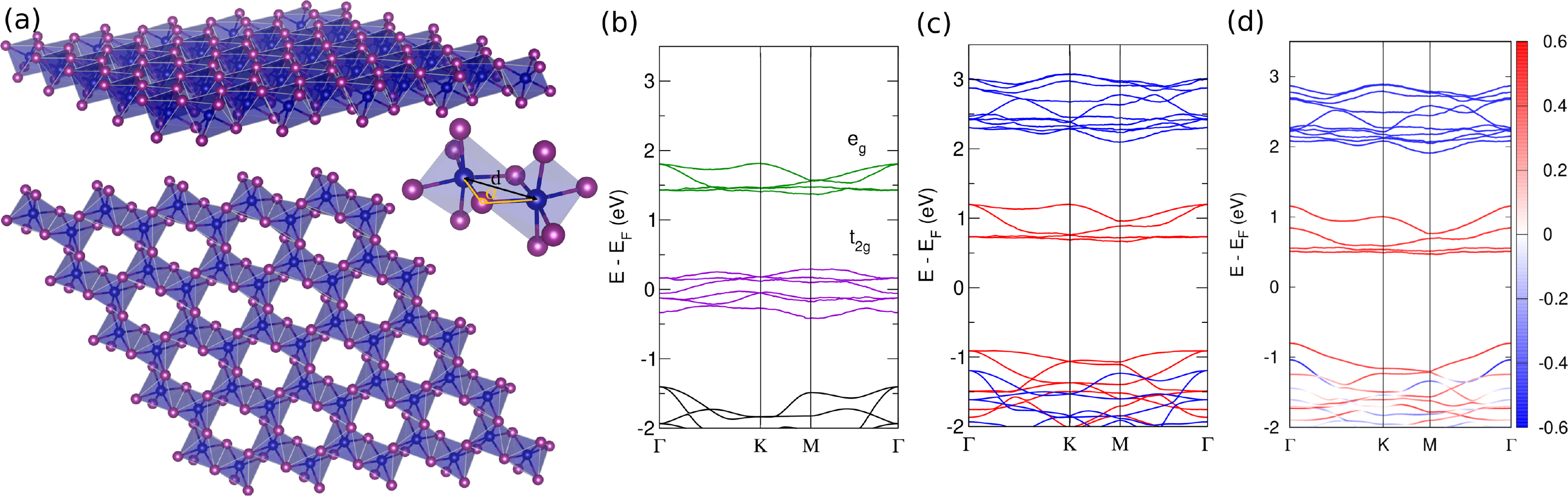}
	\caption{\label{FIG1} {\bf Atomic and electronic structure of CrI$_3$.} (a) Atomic structure of monolayer CrX$_3$. The Cr atoms (in blue) form a honeycomb structure and the ligands (X, in violet) arrange forming an octahedral frame around them. The small inset shows the unit cell, where $d$ and $\alpha$ stands for the Cr-Cr distance and the Cr-X-Cr angle respectively. (b) Band structure of CrI$_3$ in absence of magnetic exchange interactions. The $d$-orbitals split due to the octahedral crystal field induced by the ligands. (c) Spin-polarized DFT+U+J band structure ($U = 3$ and $J = 0.5$). The presence of magnetic exchange interactions spin split the half-filled $t_{2g}$ bands inducing a metal-insulator transition and a local magnetic moment $m_{Cr} = 3$ $\mu_{\rm B}$. (d) Idem with SOC.}
\end{figure*}

{\section{General background on two-dimensional magnetism}}
Many rigorous results regarding the existence, or lack thereof, of long range magnetic order in two dimensional systems are linked to model Hamiltonians, such as the Ising and the Heisenberg model. A central assumption of the theory is that the low energy magnetic properties of  insulating magnets can be described in terms of the spin Hamiltonian that can be written down as:
\begin{equation}
{\cal H}= \sum_{n,n',a,b} g_{a,b}(n,n') S_a(\vec{r}_n) S_b(\vec{r}_n)
\label{EQN1}
\end{equation}
where $n,n'$ run over the sites of a given lattice, and $a,b$ run over the 3 Cartesian indeces, $S_a(\vec{r}_n)$ is the $a$ component of the  spin operator in site $n$. This general formula includes single-ion anisotropies (when $n=n'$),   Heisenberg coupling (when $g_{a,b}(n,n')$ is proportional to the unit matrix in the $a,b$ index), Ising coupling when for $g_{a,b}(n,n')$ only the $a=b=z$ component is non zero, Dzialoshinskii-Moriya (DM) interactions, anisotropic exchange, Kitaev exchange.  Of course,  four-spin terms of the form $S_a(1) S_b(2) S_c(3) S_d(4)$ are possible in principle but are not included in eq. (\ref{EQN1}).

For a system governed by eq. (\ref{EQN1})  where all the  non-isotropic interactions are removed, Mermin and Wagner \cite{Mermin1966} demonstrated the absence of magnetic order in one- and two-dimensions at finite temperatures, previously conjectured by the spin wave theory\cite{Kranendonk1958}.
The demonstration ruled out the possibility of spontaneous magnetization at $T \neq 0$ in isotropic two-dimensional systems, however the divergences found in the spin-wave theory at finite temperatures could be lifted by introducing anisotropy. Magnetic anisotropy originates from relativistic effects such as the dipole-dipole interactions and spin-orbit coupling  that leads to magnetocrystalline anisotropy and anisotropic exchange couplings. These effects break spin rotational invariance and are fundamental to explain the existence of ferromagnetism in two dimensions.

The presence and the symmetry of magnetic anisotropy is decisively important for the properties of two-dimensional magnets. For isotropic Heisenberg ferro- or antiferromagnets (no anisotropy) at any finite temperatures $T$ the spin-spin correlation functions exponentially decay at large distances, but with the correlation length $\xi$ which is exponentially large if $T$ is much lower than a typical energy of exchange interactions $J$.\cite{Chakravarty1989} For the easy-axis anisotropy, or for the case of layered isotropic magnets with small interlayer exchange interactions, there is a phase transition to a magnetically ordered state but with transition temperatures $T_m$ much smaller than $J$, due to a strong transverse spin fluctuation; the temperature range between $T_m$ and $J$ can be described by introducing a short-range order parameter associated to local magnetic bonds.\cite{IrkhinKatsnelson1999} The Curie (or Neel) temperature is strongly suppressed in comparison with its mean-field value, by a factor $ln \Lambda$ where $\Lambda$ is the ratio of intralayer exchange interactions to either interlayer exchange interaction or to the magnetic anisotropy energy. A similar situation takes place for the Heisenberg magnet with dipole-dipole interactions.\cite{Grechnev2005} For the case of easy-plane magnetocrystalline anisotropy, there is a phase transition of Kosterlitz-Thouless type\cite{KosterlitzThouless1973}, between the states with power-law (at lower temperatures) and exponential (at higher temperatures) decay of the correlation function at large distances\cite{Hikami1980}.

Strong suppression of Curie temperature is a serious problem from the point of view of potential applications of 2D ferromagnets; this is a "revenge" of Mernin-Wagner theorem for being circumvented by weak anisotropy or interlayer exchange! The decisive role of magnetic anisotropy whatever small is the other important general feature of two-dimensional magnets.

\smallskip

{\section{Consensual aspects of chromium trihalides}} 
{\bf\em \color{blue} Electronic structure.-} CrX$_3$, with X = Cl, Br, I, belongs to the family of binary transition metal halides. In the monolayer, transition metal atoms arrange in a honeycomb structure with edge sharing octahedral coordination (Fig.\ref{FIG1}(a)). 
The oxidation state of Cr in these compounds is expected to be $+3$, with electronic configuration $[{\rm Ar}]4s^03d^3$. Therefore, from Hund's rules, we expect that Cr$^{3+}$ have $S=3/2$.  The octahedral ligand and crystal fields, split the $d$-orbitals of the Cr atoms in two subsets, namely, the $t_{2g} = \{d_{xy},d_{xz},d_{yz}\}$ and $e_g = \{d_{x^2-y^2},d_{z^2}\}$.  

For spin polarized system, a full  spin-split $t_{2g}$ band is expected, separated from the $e_g$ bands. Therefore,  ferromagnetic CrX$_3$ should be insulators according to band theory. In contrast, for a spin unpolarized solution,  we have a half-full $t_{2g}$ manifold (3 electrons in 3 spin degenerate orbitals), and band theory predicts that CrX$_3$ is a conductor (see figure (Fig. \ref{FIG1}(b))). Given that experiments show that Cr$X_3$ remains insulating in the non-magnetic state, band theory fails to describe the non-magnetic state and it can be said that Cr trihalides are correlated insulators. Fig. \ref{FIG1}(c) shows the band structure in presence of magnetic exchange interactions. \footnote{The calculation is performed using norm-conserving pseudopotentials with an energy cutoff of 40 Ry, and a $\Gamma$-centered $8\times8\times1$ k-point mesh. We use the DFT+U+J scheme within the Quantum-Espresso {\it ab-initio} package, and the Perdew-Burke-Ernzerhof (PBE) approximation for the exchange-correlation functional. For $U$ and $J$, we use similar values than the ones reported previously, namely $U = 3$ eV and $J = 0.5$ eV.\cite{LadoRossier2017}} 

{\bf\em{\color{blue} Cr-Cr intralayer exchange.-}} Density functional calculations correctly predict that intralayer exchange coupling between neighboring Cr atoms is ferromagnetic. Given their insulating nature spin-spin interactions in chromium trihalides is originated by superexchange mechanisms.\cite{LadoRossier2017,KashinRudenko2020} The computed Cr-Cr distances ($d$) and Cr-X-Cr angles ($\alpha$) (as shown in the inset of Fig. \ref{FIG1}) are summarized in Table \ref{TAB1}. 
The distance between Cr atoms decreases from I to Cl, due to the reduction of the halide atomic radius. In contrast, the angle formed by Cr-X-Cr remains almost the same for all the trihalides: 95-97$^\circ$. Ferromagnetic superexchange is expected for isovalent transition metal spins with  superexchage pathways forming  $\sim 90\deg$ angles, according to the Goodenough-Kanamori rules.
 
{\bf\em \color{blue} SOC and Magnetic anisotropy.-} Spin orbit coupling (SOC) is a main contributor to the all important magnetic anisotropy in Cr$X_3$. In Fig. \ref{FIG1}(d), we show the effect  of SOC on the band structure of monolayer CrI$_3$. While the minority $t_{2g}$ and $e_g$ bands (in blue) are weakly affected, the valence bands, with dominant contributions from the iodine $p$ orbitals\cite{LadoRossier2017} and the majority $e_g$ bands (in red) show a strong symmetry breaking and band splitting at the high-symmetry $\Gamma$-points. This already suggests that the iodine spin orbit coupling plays an important role.

DFT calculations\cite{LadoRossier2017} for CrI$_3$ show that the dominant contribution to the magnetic anisotropy energy arises from the iodine SOC.  This is expected as the single ion anisotropy  zero field splitting, of a $d^3$ cation in an octahedral environment vanishes, when only the cation SOC is considered.  In contrast, SOC from the ligand contributes to the single ion anisotropy of the Cr cation  as reported by Kim {\it et al}.\cite{KimPark2020}       

{\bf\em \color{blue} Interlayer exchange.-} In bulk,  the interlayer exchange is ferromagnetic for  CrI$_3$ and  CrBr$_3$ and antiferromagnetic   for   CrCl$_3$.   However, interlayer exchange is found to be different for thin films CrX$_3$ than bulk.  Both MOKE \cite{HuangXu2017}, and tunneling\cite{KleinJarillo2018,SongXu2018,WangMorpurgo2018},  showed that interlayer exchange is antiferromagnetic in CrI$_3$ bilayers. This is very convenient, as it permits to induce a metamagnetic transition with a magnetic field. Interlayer exchange for  CrCl$_3$ bilayers\cite{KleinJarillo2019} and thin films up to a few nanometers\cite{Serri2020} is also stronger than in bulk, although with the same sign. Therefore,  it is clear that interlayer exchange is different for atomically thin few layer systems and bulk.

CrX$_3$ planes present two different types of interlayer stacking, that leads to two different crystal structures, monoclinic and rhombohedral.
Bulk chromium trihalides display a structural transition between these two phases at temperatures way above their Curie points. In the case of CrI$_3$ and CrCl$_3$, a phase transition from monoclinic to rhombohedral crystal structure has been observed below room temperature ($T \sim 210$ and $240$ K respectively). This crystal phase transition is usually accompanied by a change in the magnetic susceptibility or an increase of the ferromagnetic signal.\cite{McGuireSales2015,KleinJarillo2019} 

The unexpected finding of  interlayer coupling for CrI$_3$ and CrCl$_3$ different in thin films and in bulk has led to explore the interplay between interlayer exchange and stacking, and to conjecture that the structural transition between monoclinic and rhombohedral may not be present.  \cite{SorianoRossier2019,JiangJi2019,SivadasXiao2018,JangHan2018}. Unsurprisingly,  all these DFT works show that interlayer exchange is indeed different for the two stackings, with a tendency to promote AF layer in the monoclinic phase of CrI$_3$. Detailed analysis of the interaction between Cr orbitals in different layers has shown that $t_{2g}$-$t_{2g}$ interactions are more favored in the monoclinic stacking leading to an AFM interlayer coupling\cite{SivadasXiao2018}. In contrast, in the rombohedral stacking, the $t_{2g}$-$e_g$ interactions are favoured leading to a FM exchange coupling. Recently, a experimental work by Ubrig {\it et al}\cite{UbrigGibertini2019} has proven that for CrI$_3$ this phase transition is thickness dependent, with monoclinic crystal structure even at low temperature.

\begin{table}[t]
	\caption{Cr-Cr distance ($d$), Cr-X-Cr angle ($\alpha$), Curie temperature ($T_C$) of single-layer (and bulk), and type of anisotropy of monolayer CrX$_3$. [$^*$This temperature corresponds to the bilayer.]}
	\label{TAB1}
	\begin{tabular}{lcccc}
		\hline
		& \multicolumn{1}{l}{\textbf{d$_{\rm \bf Cr-Cr}$(Å)}} & \multicolumn{1}{l}{\textbf{a(º)}} & \multicolumn{1}{l}{\textbf{T$_{\rm C}$(K)}} & \multicolumn{1}{l}{\textbf{Type of anisotropy}} \\ \hline
		\textbf{CrI$_{\bf 3}$}  & 4.026                                   & 97.5                               & 46 (61)                             & Easy axis (z)                                   \\
		\textbf{CrBr$_{\bf 3}$} & 3.722                                   & 94.9                               & 27 (37)                             & Easy axis (z)                                   \\
		\textbf{CrCl$_{\bf 3}$} & 3.491                                   & 95.5                               & 16* (17)                            & Easy plane (xy)                                 \\ \hline
	\end{tabular}
\end{table}

\begin{figure}
	\centering
	\includegraphics[clip=true, width=0.9\columnwidth] {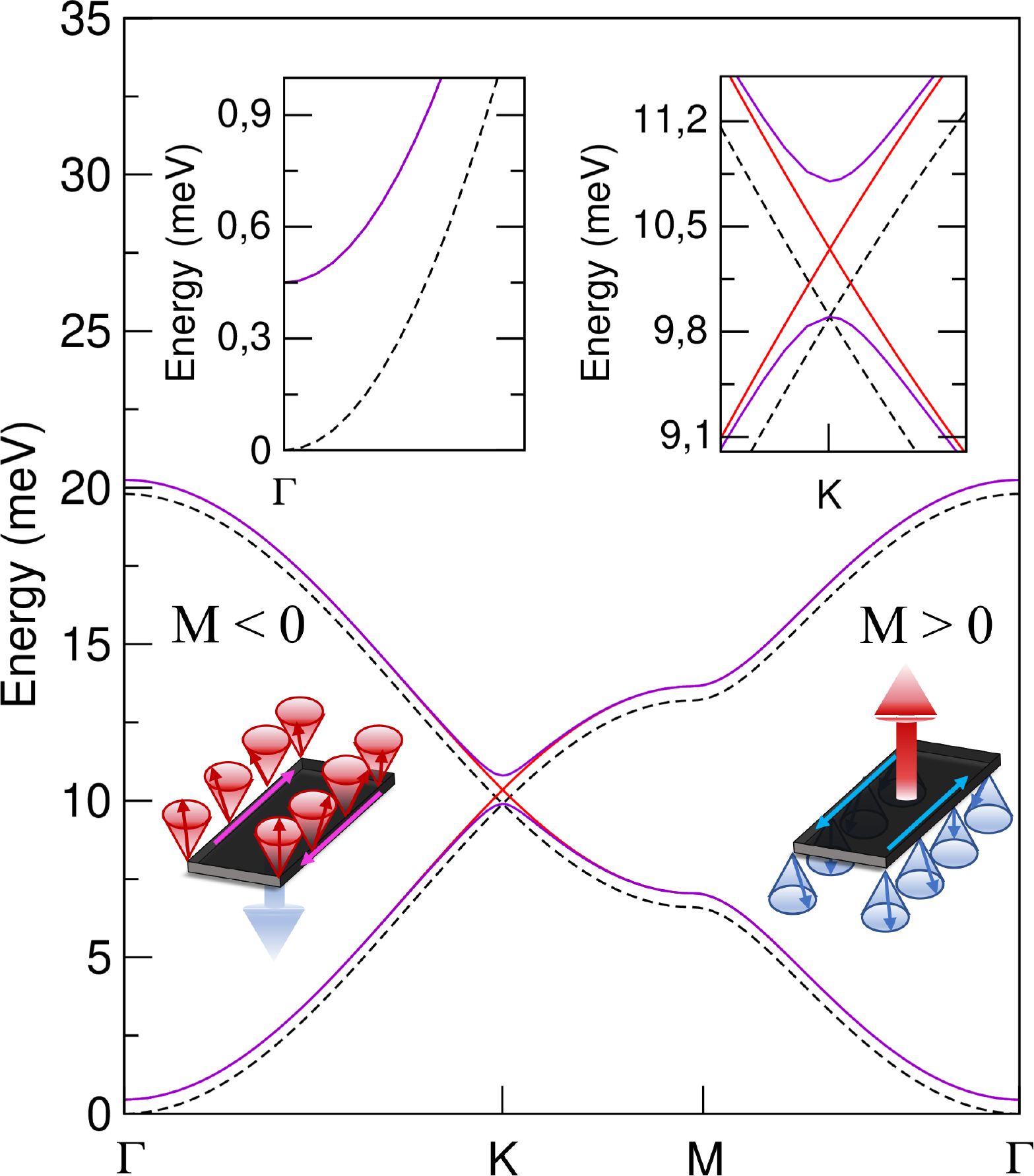}
	\caption{\label{FIG2} {\bf Spin wave dispersion of a honeycomb CrI$_3$ layer.} The spin-wave dispersion shows similarities with the tight-binding model in a honeycomb lattice. In absence of any type of anisotropy (dashed line), it shows parabolic bands at $\Gamma$ and a Dirac-like dispersion at $K$. The presence of single-ion and/or XXZ-like anisotropy (red line), a gap opens at $\Gamma$ (top-left inset). An additional topological gap can be opened at $K$ (violet line) by including second-neighbor Dzyaloshinskii-Moriya or Kitaev interactions (top-right inset). This topological phase hots counterpropagating gapless edge topological magnons as those shown on the left and right sides of the figure for the two magnetic ground states: $M > 0$, and $M < 0$ respectively.}
\end{figure}

{\bf\em \color{blue} Magnons.-} Given that Cr atoms form a honeycomb lattice, the magnon dispersion has two branches, acoustic and optical. Linear spin wave theory for an off-plane ferromagnet with first Heisenberg neighbour interactions is mathematically isomorphic to the one orbital tight-binding model for graphene electrons. Thus, the acoustic and optical branches meet at the corners of the Brillouin zone, forming Dirac cones  and a quadratic dispersion at the $\Gamma$ point (Fig. \ref{FIG2}).  When magnetic anisotropy is included\cite{LadoRossier2017} in the form of either single ion uniaxial anisotropy $A$ or  anisotropic $XXZ$ super-exchange $J_z$, a band-gap  $\Delta=2AS + 3J_z S$, with $S=3/2$ opens at the $\Gamma$ point, that is essential to  stabilise magnetism (see top-left inset in Fig. \ref{FIG2}).

The optimal tool to probe magnon energy dispersion is inelastic neutron scattering. Unfortunately, this is only viable in bulk samples. There, a gap at the Dirac points \cite{ChenDai2018} is found in CrI$_3$, and it is attributed to a topological origin. This is confirmed by a first-principles calculation of the magnon dispersion based on the itinerant fermion description\cite{CostaRossier2020}. In a spin model picture, two different anisotropic spin couplings are predicted to open a topological gap in a honeycomb ferromagnet:  second neighbour DM ($D'$) \cite{Owerre2016} and  first neighbour Kitaev ($K$)\cite{AguileraNunez2020}.  Further insight on the spin couplings can be  gained from ferromagnetic resonance, that  permits to infer the strength of Heisenberg and Kitaev interactions from the resonance field of bulk CrI$_3$.\cite{LeeHammel2020}

Experimental techniques to probe magnons in monolayer, or few-layer, samples are limited to Raman\cite{JinHe2018} or inelastic tunneling spectroscopy (IETS).\cite{KleinJarillo2018,KimTsen2019} Any of these 2D techniques has access to the magnon dispersion, but allows to explore the magnon frequencies (Raman), DOS and gap via $d^2I/dV^2$ (IETS). Most of the experimental methods infer a magnon gap smaller than 1 meV, much larger than MAE in conventional ferromagnets, and in line with DFT calculations.
 
{\bf\em \color{blue} Critical temperature.-}  In the case of off-plane easy axis 2D CrI$_3$ and CrBr$_3$, the Curie temperature is governed by two main  scales,  the MAE and the spin stiffness, controlled by exchange. 
In the Ising limit, where MAE is much larger than the spin stifness, magnetic order is destroyed by proliferation of domains with opposite magnetization. In the opposite Heisenberg limit, magnetic order is depleted by proliferation of magnons with long wavelength.  In both cases there are analytical expressions to describe these phenomena. An interpolation formula has been recently proposed\cite{TorelliOlsen2019} that seems to work pretty well in the case of CrI$_3$. 

{\bf\em \color{blue} Magneto-optical properties.-} Magneto-optical Kerr effect (MOKE) and magnetic circular dichroism (MCD) measurements permit to probe magnetic order down to the monolayer limit, and made possible the discovery of ferromagnetism in 2D materials, such as CrI$_3$ and CrGeTe$_3$. The coupling between the optical response and magnetism is necessarily associated to spin orbit coupling. First  principles calculation show that the strength of the MOKE is governed by the spin orbit coupling of the ligand\cite{MolinaRossier2019}. These calculations, that include a GW description of the quasiparticles and solve the Bethe-Salpeter equation,  also show a very strong excitonic effect in the 3 chromium trihalides, found earlier for CrI$_3$.\cite{WuLouie2019} 

In contrast to transition metal dichalcogenides, where excitons show mostly Wannier-Mott-like behaviour (large exciton radii), excitons in  chromium trihalides are closer to the Frenkel-like picture. Interestingly, the exciton radii increases when going from CrCl$_3$ to CrI$_3$. This seems to be related to the strong Cr-ligand hybridization in the valence bands of CrI$_3$. Despite the recent advances on this field, the topic is still in its infancy. Therefore, new and very interesting phenomena might be reported in the near future.       

\section{Open questions in the theory of chromium trihalides} 

{\bf\em \color{blue} Intra- and interlayer exchange interactions.-} 
 The first open question is the nature of both intralayer and interlayer Heisenberg exchange interactions. Intralayer exchange is clearly ferromagnetic for  $X=Cl,Br,I$, and the almost 90 degree angle formed by the Cr-ligand-Cr superexchange pathway fits the Goodenough-Kanamori scenario for ferromagnetic super-exchange. However, a quantitative understanding of the role played by local Coulomb interactions, orbital environment, and ligand substitution on the exchange interaction deserve further attention. In this regard, different theoretical approaches combining total energy (DFT+U) calculations\cite{LadoRossier2017,TorelliOlsen2019,ZhangLam2015}, multiconfigurational\cite{PizzocheroYazyev2020} and Green's function methods\cite{BesbesSolovyev2019,KashinRudenko2020} have been used so far to shed some light on these questions. In Table \ref{TAB2}, we summarize the theoretical and experimental exchange values ($J$) reported so far. 

\begin{table*}[t]
	\caption{{\bf Summary of the values reported so far for the different  exchange parameters in CrX$_3$.} The values are obtained experimentally ($^{\color{red} \rm E}$) and theoretically ($^{\color{red} \rm T}$) for different type of samples, namely, monolayers(ML), few-layers(FL) and bulk. $B_\parallel^c$ and $B_\perp^c$ stand for the critical fields in the parallel and perpendicular directions with respect to the layers. $J_z$ is the anisotropic exchange. $J_i$ are the Heisenberg exchange interactions for first, second and third neighbors. $D'$ and $K$ correspond to the second neighbor and Kitaev interactions leading to a non-trivial gap opening at the $K$-point of the magnon spectrum. $A$ is the single-ion anisotropy.} 
		\label{TAB2}
	\resizebox{\textwidth}{!}{
	\begin{tabular}{llllllllll}
		\hline
		& \bf B$_{\bf \parallel}^{\bf c}$({\bf T})                                                                     & \bf B$_{\bf \perp}^{\bf c}$(T)                                                                                                              & \bf MAE(meV)                                                                  & \bf J$_{\bf z}$   (meV)                                                                                                                & \bf J$_{\bf 1}$   (meV)                                                                                                                                                                                                                       & \bf J$_{\bf 2}$   (meV)                                                                                                             & \bf J$_{\bf 3}$   (meV)                                                                                           & \bf D’/K (meV)         & \bf A (meV)                                                                                                                                                                                      \\ \hline
		\bf CrI$_{\bf 3}$  & \begin{tabular}[c]{@{}l@{}}6.5{\color{red}$^{\rm E}$}\cite{KimTsen2019,WangMorpurgo2018}{[}FL{]}\\    3.8{\color{red}$^{\rm E}$}\cite{SongXu2018}{[}BL{]}\end{tabular} & \begin{tabular}[c]{@{}l@{}}2.0{\color{red}$^{\rm E}$}\cite{KimTsen2019}{[}FL{]}\\    1.8{\color{red}$^{\rm E}$}\cite{KleinJarillo2018,SongXu2018}{[}FL{]}\\    0.6{\color{red}$^{\rm E}$}\cite{SongXu2018}{[}BL{]}\\    0.85{\color{red}$^{\rm E}$}\cite{KleinJarillo2018}{[}BL{]}\end{tabular} & \begin{tabular}[c]{@{}l@{}}0.65{\color{red}$^{\rm T}$}\cite{LadoRossier2017}{[}ML{]}\\    0.68{\color{red}$^{\rm T}$}\cite{ZhangLam2015}{[}ML{]}\end{tabular} & \begin{tabular}[c]{@{}l@{}}2.38{\color{red}$^{\rm E}$}\cite{KimTsen2019}{[}FL{]}\\    0.27{\color{red}$^{\rm E}$}\cite{LeeHammel2020}{[}Bulk{]}\\    0.022{\color{red}$^{\rm T}$}\cite{TorelliOlsen2019}{[}ML{]}\\    0.09{\color{red}$^{\rm T}$}\cite{LadoRossier2017}{[}ML{]}\end{tabular} & \begin{tabular}[c]{@{}l@{}}2.29{\color{red}$^{\rm E}$}\cite{KimTsen2019}{[}FL{]}\\    2.01{\color{red}$^{\rm E}$}\cite{ChenDai2018}{[}Bulk{]}\\    0.20{\color{red}$^{\rm E}$}\cite{LeeHammel2020}{[}Bulk{]}\\    3.24{\color{red}$^{\rm T}$}\cite{TorelliOlsen2019}{[}ML{]}\\    2.2{\color{red}$^{\rm T}$}\cite{LadoRossier2017}{[}ML{]}\\    1.44{\color{red}$^{\rm T}$}\cite{BesbesSolovyev2019}{[}ML{]}\\    1.0{\color{red}$^{\rm T}$}\cite{KashinRudenko2020}{[}ML{]}\\    2.86{\color{red}$^{\rm T}$}\cite{ZhangLam2015}{[}ML{]}\\    2.29{\color{red}$^{\rm T}$}\cite{XuBellaiche2018}{[}ML{]}\end{tabular} & \begin{tabular}[c]{@{}l@{}}0.16{\color{red}$^{\rm E}$}\cite{ChenDai2018}{[}Bulk{]}\\    0.56{\color{red}$^{\rm T}$}\cite{TorelliOlsen2019}{[}ML{]}\\    0.4{\color{red}$^{\rm T}$}\cite{KashinRudenko2020}{[}ML{]}\\    0.63{\color{red}$^{\rm T}$}\cite{ZhangLam2015}{[}ML{]}\end{tabular} & \begin{tabular}[c]{@{}l@{}}-0.1{\color{red}$^{\rm E}$}\cite{ChenDai2018}{[}Bulk{]}\\    0.001{\color{red}$^{\rm T}$}\cite{TorelliOlsen2019}{[}ML{]}\\    -0.15{\color{red}$^{\rm T}$}\cite{ZhangLam2015}{[}ML{]}\end{tabular} &
		\begin{tabular}[c]{@{}l@{}}5.2{\color{red}$^{\rm E}$}\cite{LeeHammel2020}{[}Bulk{]}\\    0.08{\color{red}$^{\rm T}$}\cite{BesbesSolovyev2019}{[}ML{]}\\    0.85{\color{red}$^{\rm T}$}\cite{ChenGao2019}{[}ML{]}\\ 0.3{\color{red}$^{\rm E}$}\cite{ChenDai2018}{[}Bulk{]}\end{tabular} & \begin{tabular}[c]{@{}l@{}}0.22{\color{red}$^{\rm E}$}\cite{ChenDai2018}{[}FL{]}\\    0.056{\color{red}$^{\rm T}$}\cite{TorelliOlsen2019}{[}ML{]}\\    0.1{\color{red}$^{\rm E}$}\cite{BesbesSolovyev2019}{[}ML{]}\end{tabular}  \\
 \hline
		\bf CrBr$_{\bf 3}$ & 0.4{\color{red}$^{\rm E}$}\cite{KimTsen2019}{[}FL{]}                                                                & \textless 0.01{\color{red}$^{\rm E}$}\cite{KimTsen2019}{[}FL{]}                                                                                              & 0.18{\color{red}$^{\rm T}$}\cite{ZhangLam2015}{[}ML{]}                                                              & 1.58{\color{red}$^{\rm E}$}\cite{KimTsen2019}{[}FL{]}                                                                                                           & \begin{tabular}[c]{@{}l@{}}1.56{\color{red}$^{\rm E}$}\cite{KimTsen2019}{[}FL{]}\\    2.6{\color{red}$^{\rm T}$}\cite{ZhangLam2015}{[}ML{]}\end{tabular}                                                                                                                                                      & 0.38{\color{red}$^{\rm T}$}\cite{ZhangLam2015}{[}ML{]}                                                                                                        & -0.15{\color{red}$^{\rm T}$}\cite{ZhangLam2015}{[}ML{]}                                                                                     &                  &                                                                                                                                                                                                      \\
		\hline
		\bf CrCl$_{\bf 3}$ & 2.0{\color{red}$^{\rm E}$}\cite{KimTsen2019}{[}FL{]}                                                                & \begin{tabular}[c]{@{}l@{}}2.4{\color{red}$^{\rm E}$}\cite{KimTsen2019}{[}FL{]}\\    0.85{\color{red}$^{\rm E}$}\cite{KleinJarillo2019}{[}BL{]}\\    1.6{\color{red}$^{\rm E}$}\cite{KleinJarillo2019}{[}FL{]}\end{tabular}                        & 0.03{\color{red}$^{\rm T}$}\cite{ZhangLam2015}{[}ML{]}                                                              & 0.91{\color{red}$^{\rm E}$}\cite{KimTsen2019}{[}FL{]}                                                                                                           & \begin{tabular}[c]{@{}l@{}}0.92{\color{red}$^{\rm E}$}\cite{KimTsen2019}{[}FL{]}\\    1.92{\color{red}$^{\rm T}$}\cite{ZhangLam2015}{[}ML{]}\end{tabular}                                                                                                                                                     & 0.23{\color{red}$^{\rm T}$}\cite{ZhangLam2015}{[}ML{]}                                                                                                        & -0.13{\color{red}$^{\rm T}$}\cite{ZhangLam2015}{[}ML{]}                                                                                     &                  &                                                                                                                                                                                                    \\ \hline
	\end{tabular}
}
\end{table*}

The origin of the anomalous interlayer exchange in CrX$_3$ thin films, discussed above, is not completely understood. It is clear that super-exchange occurs through two ligands, so that simple Goodenough-Kanamori rules no longer apply. Both the experimental results and the DFT calculations suggest that there are at least two  competing interlayer exchange interactions with different signs in CrX$_3$ multilayers, and whose nature is yet to be determined. Their competition   should permit to understand the dependence of the interlayer exchange on the thickness of the sample. Here, two different scenarios have been proposed. In the first one,\cite{SivadasXiao2018,JangHan2018,SorianoRossier2019,LeiMacDonald2019,JiangJi2019} the interlayer exchange would be different because thin films would not undergo the structural stacking transition that happens in bulk well above the Curie temperature. Of course, this scenario would open a second question, namely, why there is no structural transition in thin films. In the second scenario, the difference between interlayer exchange in thin films would arise from  a larger interlayer distance in few layer systems, due to a smaller contribution of the long range Van der Waals interaction.  

Modelling interlayer exchange is particularly challenging for DFT for  several reasons. First, it is a small quantity that is not much higher than the accuracy of the codes.  Second, there are competing mechanisms that contribute, whose  relative weight depends strongly on the distance. Third,  local density functionals perform poorly to describe the tails of wave functions, that control interlayer exchange.  As a result, it does  not come as a surprise the very large dispersion of values and signs of the interlayer exchange computed with different functionals,\cite{JiangJi2019} although all of them support the trend by which the interlayer exchange tends to be AF for CrI$_3$ bilayers  in the monoclinic phase.

A number of experiments using a variety of techniques, such as second harmonic generation,\cite{SunWu2019} Raman,\cite{McCrearyHight2019,ZhangHuang2020} NV magnetometry,\cite{ThielMaletinsky2019}  spin resonance,  have addressed this matter. The theoretical picture based on the interplay between stacking and magnetic order in few-layer CrI$_3$ has been corroborated by different experiments. For instance, room temperature exfoliated thin layers of CrI$_3$ have shown to turn from layered AFM to layered FM after accidental puncture of the sample at 7 K in NV-magnetometry.\cite{ThielMaletinsky2019} Similarly, Chen et al.\cite{ChenGao2019} have demonstrated that stacking plays a crucial role on the interlayer exchange in thin samples of CrBr$_3$ grown by molecular beam epitaxy. They show how different stacking patterns in bilayer structures lead to different interlayer magnetic order. The presence of a monoclinic phase has also been verified using second harmonic generation, which is sensitive to crystal symmetries,\cite{SunWu2019} and polarization resolved Raman spectroscopy.\cite{UbrigGibertini2019} Nonetheless, the recent observation of an enhancement of interlayer AFM exchange in ultrathin samples of CrCl$_3$ with respect to bulk samples,\cite{KleinJarillo2019} may point in the direction that van der Waals interactions are playing an important role in the mechanism of interlayer exchange in chromium trihalides.     

{\bf\em \color{blue} Origin of magnetic anisotropy.-} The second big open question in our understanding of Cr trihalides is the nature of the anisotropic interactions. These anisotropic terms control two very important quantities, the magnetic anisotropy energy (MAE) and the concomitant gap of the acoustic branch of the spin waves, and the gap at the Dirac points, that may be of topological origin.\cite{ChenDai2018,LeeHammel2020} All calculations agree on the fact that magnetic anisotropy is mostly determined by the spin orbit of the ligand\cite{ZhangLam2015,LadoRossier2017,PizzocheroYazyev2020,XuBellaiche2018}. However,  there is no consensus on the value of the various anisotropic terms that are permitted by symmetry in the Hamiltonian,  such as the   single ion anisotropy ($A$),  anisotropic exchange ($J_z$), Kitaev exchange ($K$) and second neighbor DM exchange ($D'$) (Table \ref{TAB2}).   

Calculations indicate that the important gap at the $\Gamma$-point in the spin-wave spectrum (Fig. \ref{FIG2}), that prevents the infrared magnon catastrophe,  depends on the spin orbit of the ligand\cite{LadoRossier2017,CostaRossier2020}. However, both the single ion anisotropy and the anisotropic exchange can also contribute.  In principle, the octahedral symmetry of the Cr crystal field gives a vanishing single ion uniaxial anisotropy, when the spin orbit coupling of the ligand is ignored.  This, together with the strong dependence of MAE on the spin orbit of iodine,  led one of us to propose that anisotropic exchange had to arise from super-exchange. However, recent multi-reference calculations\cite{PizzocheroYazyev2020} and spin resonance experiments indicate otherwise.  In addition,  it has been recently 
proposed that  spin orbit coupling of the ligand can yield a significant contribution to the single ion anisotropy\cite{KimPark2020}.This issue deserves further attention. 

\section{Heterostructures  and applications}

The outstanding  physical phenomena that constitute the core of spintronics,  such as  Giant Magneto Resistance,  Tunnel Magneto Resistance (TMR)  and Spin transfer Torque, occur in artificial structures that integrate layers of magnetic and non magnetic materials. Therefore, the discovery of magnetic 2D crystals, and the relative simplicity of their integration in vertical Van der Waals heterostructures that do not need to address the problem of epitaxial growth,  have a huge potential to open new venues in spintronics. There are of course challenges that need to be addressed.  Room temperature feromagnetism has only been reported for a few  materials, definitely not for the Cr trihalides. Yet,  Cr$X_3$ can be used in  possible device concepts that work at low temperature and can be later implemented with room temperature ferromagnets.

\begin{figure*}
	\centering
	\includegraphics[clip=true, width=0.8\textwidth] {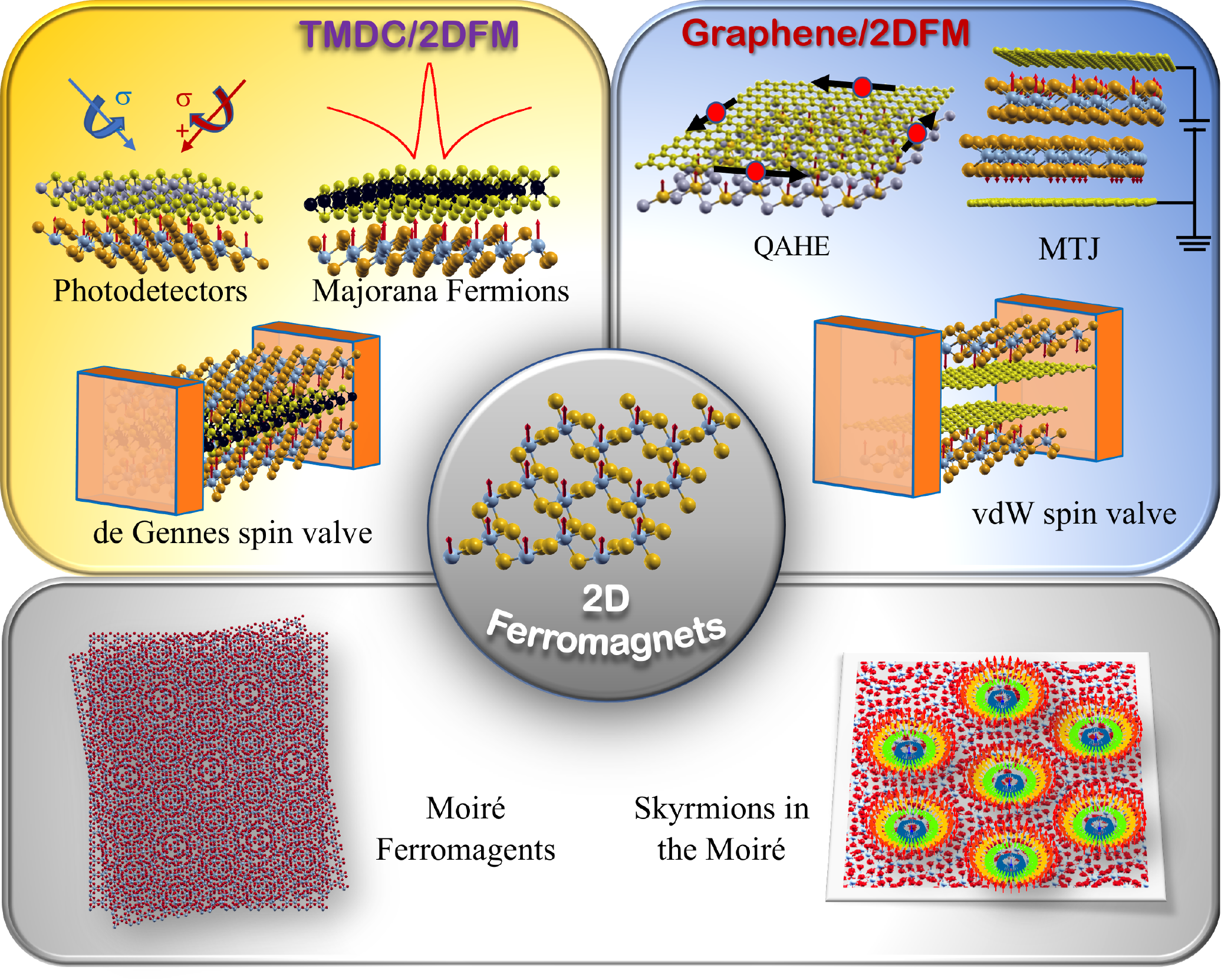}
	\caption{\label{FIG3} {\bf Applications of 2D ferromagnets in van der Waals heterostructures and twisted lattices.} }
\end{figure*}

{\bf\em \color{blue} Spin filter tunnel junctions.-} One of the first novel device concepts made possible using atomically thin chromium trihalides are the so called spin filter magnetic tunnel junctions (MTJ). Unlike conventional  MTJ,  that feature a dull insulating tunnel barrier sandwiched between two conducting ferromagnets,  spin filter MTJ have tunnel barriers with two ferromagnetic insulators whose magnetization can be oriented independently,  making possible a spin-valve action.  The figure of merit in these systems is the TMR, defined as the  change of resistance  between the paralell and anti-parallel states of the two ferromagnets. Taking advantage of the AF Interlayer exchange that CrI$_3$ has for thin films, several groups reported giant TMR  on graphite/CrI$_3$/graphite MTJ.\cite{KleinJarillo2018,WangMorpurgo2018,SongXu2018} This magnetoresistance increases for thicker magnetic tunnel barriers up to 19000\%, this is around one order of magnitude higher than those based on MgO. However,  large TMR is observed only at low temperatures and is larger at higher bias voltage\cite{SongXu2018},  a behaviour opposite to conventional MTJ that deserves to be studied from the theory side.

{\bf\em \color{blue}  Electric field control.-} The insulating nature of CrX$_3$ opens the gate for electrical manipulation of magnetism.\cite{JiangShan2018,JiangMak2018,HuangXu2018} Dual-gated bilayer CrI$_3$ devices have shown to undergo a AFM-FM transition for electron doping of the order of $n \sim 2\times10^{13}$ cm$^{-2}$. This is in agreement with recent theoretical calculations, predicting the formation of magnetic polarons for electron-doped bilayer CrI$_3$.\cite{SorianoKatsnelson2020}

{\bf\em \color{blue}  Magnetic photogalvanic effect.-} In non-centrosymmetric materials, strong light irradiation transforms into a dc current by the bulk photogalvanic effect. The magnetic photogalvanic effect occurs in magnetic materials with strong spin-orbit coupling. Zhang \emph{et al} have recently predicted this effect in bilayer CrI$_3$.   \cite{ZHangYan2019} A photocurrent can be generated when the relative orientation of the magnetization in each layer is antiparallel. This, together with SOC breaks momentum-inversion symmetry ($E(-\vec{k}) \neq E(\vec{k})$). The dc current cancels for parallel orientation. 

{\bf\em \color{blue} Spin orbit torques.-} Spin orbit torques can provide another way to induce magnetic switching in CrX$_3$. This has been explored computationally by Dolui {\emph et al}\cite{DoluiNikolic2020} on bilayer-CrI$_3$/TaSe$_2$ van der Waals heterostructure. They show that an unpolarized current injected along the metallic TaSe$_2$ is spin-polarized due to the strong spin-orbit and, eventually, jump into the adjacent CrI$_3$ layer switching the magnetization without requiring any external field.    
  
{\bf\em \color{blue}  Exchange proximity effect.-} Exchange proximity effect in van der Waals heterostructures is the main ingredient in many device concepts, combining CrX$_3$ with either conductors such as graphene bilayers \cite{CardosoRossier2018},  semiconducting transition metal dichalcogenides 
\cite{ZhongXu2017,ZhongXu2020,CiorciaroImamoglu2020,SeylerXu2018,LyonsTartakovskii2020}
and even recent experiments with superconductors.\cite{KezilebiekeLiljeroth2020}

Exchange proximity effect in 2D-FM/Graphene vertical devices has been partially unexplored experimentally. From the theoretical side, some interesting proposals has been already reported to use these bilayer heterostructures as spin filters.\cite{CardosoRossier2018}  In absence of pressure, the exchange proximity effect obtained from first-principles calculations is very small, in the range of a few meV. However, the hybridization of the Dirac cones of graphene with the almost flat spin-polarized $e_g$ bands of CrI$_3$ destroy completely one of the two spin channels present in Graphene, letting one completely open for conduction. Based on this, Cardoso \emph{et al} proposed a bilayer graphene sandwiched between ferromagnetic insulator CrI$_3$. Interestingly, when adjacent ferromagnetic layers are ordered parallel with respect to each other, an open spin channel is available in both graphene layers. In contrast, when they ordered anti-parallel, the spin hybridization between graphene layers is destructive and the conduction is expected to decrease dramatically. 

Another very appealing possibilty is to turn  graphene into a Chern insulator due to magnetic proximity with CrI$_3$.  According to DFT calculations \cite{ZhangYang2018} this would happen in graphene/CrI$_3$ under compressive pressure ($1.7$ - $18.3$ GPa).  The quest of other combinations of magnetic and non-magnetic materials that lead to topological gaps is a very promising research venue.

The valley polarization of the photoluminescence  of 2D transition metal dichalcogenides (TMDs) have proven to be  affected by proximitty to thin films of  CrI$_3$ \cite{ZhongXu2017,ZhongXu2020,CiorciaroImamoglu2020,SeylerXu2018} and to CrBr$_3$\cite{LyonsTartakovskii2020}
In all these structures,  the combined effect of exchange field and intrinsic spin-orbit coupling is known to break spin-valley symmetry in TMDs leading to a different excitation spectra for left and right circularly polarized light.  This allows to use the 2D TMD to probe the magnetism of the top layer of the CrX$_3$ stack.

Heterostructures that combine ferromagnetic insulators and superconductors could induce spin triplet superconductivity, and even  topological superconductivity, with potential applications in  topological quantum computing.  Interestingly,  a first experiment studying STM transport  in CrBr$_3$/NbSe$_2$ reports the obervation of a zero bias peak consistent with a Majorana zero mode\cite{KezilebiekeLiljeroth2020}.  This matter certainly deserves further scrutiny from the theory side. 

Heterostructures combining ferromagnetic insulators, such as CrX$_3$ and atomically thin superconductors,   seem and ideal arena to test the 50 year  old  proposal by de Gennes.\cite{DeGennes1966} There, a superconducting spin valve is made by sandwiching a superconductor between two ferromagnetic insulators. When the magnets order parallel the exchange field destroys the superconducting phase and the conduction vanish, resulting in a change between a state with zero resistance, and one with rather large resistance, on account of the very small section of the superconducting 2D crystal.  The recent advances in the fabrication of van der Waals devices, together with the discovery of 2D ferromagnetic insulators and 2D superconductivity seems a low lying fruit.

{\bf\em \color{blue}  Magnetic multilayers, Moir\'es and twisted 2D magnets.-} Exchange bias, namely, the shift of the magnetization hysteresis cycle (M(H)) due to coupling to an antiferromagnet, has been recently observed in Fe$_3$GeTe$_2$/CrCl$_3$, where CrCl$_3$ plays the role of the antiferromagnet.\cite{ZhuWee2020} Exchange bias has played a very important role as a resource in applied magnetism, yet the precise microscopic mechanism are not fully understood theoretically. This is yet another area of research. 

In the case of van der Waals magnetic multilayers, the twisting angle is an additional degree of freedom that is frozen in epitaxially grown conventional metallic heterostructures. This has an enormous potential for new physical phenomena. The implications for the magnons are already being explored at the theoretical level. The emergence of novel and very fascinating physics in twisted bilayer graphene, has stimulated researchers to look into the electronic properties of twisted bilayer structures in general.  Chromium trihalides are very intriguing in this sense, since they show an additional ingredient, namely a stacking dependent interlayer magnetism. An overview on Moir\'e magnets has been worked out recently by Hejazi {\emph et al.}\cite{HejaziBalents2020}  A common interesting feature in the Moir\'e patterns of these ferromagnets is that, for certain angles, the magnons become localized with a spectrum showing flat dispersion. Also, in twisted 2D ferromagnets, inversion symmetry is broken and first neighbour DMI is no longer zero. The emergence of skyrmions in the bulk of twisted CrBr$_3$ has been already proposed by Tong {\emph et al}.\cite{TongYao2018}

\section{Conclusion and outlook}
In summary,  the exploration of the properties of monolayer and few layer family of CrX$_3$ is being an amazingly versatile and fruitful arena to explore many fascinating physical phenomena and the implementation of several device concepts. Thus, CrI$_3$ layers can be used to build spin filter tunnel junctions with record high tunnel magneto-resistance, they host topological magnons, they can induce topological gaps in graphene and perhaps even  topological superconductivity in NbSe$_2$. 

Important theory questions, such as the nature of the anisotropic spin interactions and the thickness dependence of the interlayer coupling, remain to be addressed. The exploration of structures that combine CrX$_3$ with other materials,  or even different types of CrX$_3$, is also a fertile terrain for new discoveries. Open theory questions such as  how  quantum fluctuations that prevent broken symmetry in 1D structures, such as CrX$_3$ nanotubes and nanoribbons, may become relevant if these structures are fabricated. This is something to keep in mind, given the recent report on 0D  CrI$_3$ nanoplatelets.\cite{SienaGamelin2020}

\bibliography{biblio}

\end{document}